\documentclass[manuscript]{aastex62}
\usepackage{bm}
\usepackage{epsfig}
\usepackage{graphicx}
\usepackage{color}
\usepackage{subfigure}
\usepackage{amsfonts,amssymb,amsmath}

\usepackage{color}
\usepackage{amssymb}
\usepackage{graphicx}
\usepackage{dcolumn}
\usepackage{bm}

\submitjournal{ApJ}
\shorttitle{Inertial Kinetic-Alfv\'en Turbulence}
\shortauthors{Roytershteyn et al.}

\begin{document}

\title{Numerical Study of Inertial Kinetic-Alfv\'en Turbulence}

\correspondingauthor{Vadim Roytershteyn}
\email{vroytershteyn@spacescience.org}

\author[0000-0003-1745-7587]{Vadim Roytershteyn}
\affiliation{Space Science Institute, Boulder, Colorado 80301, USA}
\author{Stanislav Boldyrev}
\affiliation{Space Science Institute, Boulder, Colorado 80301, USA}
\affiliation{Department of Physics, University of Wisconsin at Madison, Madison, WI 53706, USA}
\author{Gian~Luca Delzanno}
\affiliation{T-5 Applied Mathematics and Plasma Physics Group, Los Alamos National Laboratory, Los Alamos, NM 87545, USA}
\author{Christopher~H.~K.~Chen}
\affiliation{School of Physics and Astronomy, Queen Mary University of London, London, E1 4NS, UK}
\author{Daniel Gro\v selj}
\affiliation{Max-Planck-Institut f{\"u}r Plasmaphysik, Boltzmannstra{\ss}e 2, D-85748 Garching, Germany}
\author{Nuno F. Loureiro}
\affiliation{Plasma Science and Fusion Center, Massachusetts Institute of Technology, Cambridge, Massachusetts 02139, USA}


\date{\today}

\begin{abstract}
Recent observational and analytical studies suggested that a new regime of kinetic turbulence may exist in plasma environments with low electron beta~\cite[][]{chen_boldyrev2017}. Such a regime, termed inertial kinetic-Alfv\'en turbulence, is relevant for the solar corona, Earth's magnetosheath, and other astrophysical systems where the electron and ion plasma beta parameters satisfy the condition 
$\beta_e\ll \beta_i\lesssim 1$. In this paper we present kinetic numerical simulations that confirm existence of the iKAW regime. Specifically, the simulations demonstrate a transition at scales below electron inertial length $d_e$ when $\beta_e\ll \beta_i\lesssim 1$. Spectral slopes and other statistical properties of turbulence at sub-$d_e$ scales are consistent with the phenomenological theory of inertial kinetic-Alfv\'en turbulence proposed by~\cite{chen_boldyrev2017} and with the recent observations in the Earth's magnetosheath.   
\end{abstract}

\section{Introduction}
Observations of the solar wind show plasma fluctuations over a broad range of scales. 
These fluctuations play a role in plasma heating, particle acceleration, heat conduction, magnetic reconnection, and other processes~\cite[e.g.][]{bruno:2005,Kiyani2015}. 
Most of the currently available analytical and numerical studies have been devoted to distances relatively far from the Sun, exceeding $\sim 0.3$~AU, since such regions have been most accessible to the existing space missions. 

\begin{figure}[h]
\includegraphics[width=\columnwidth]{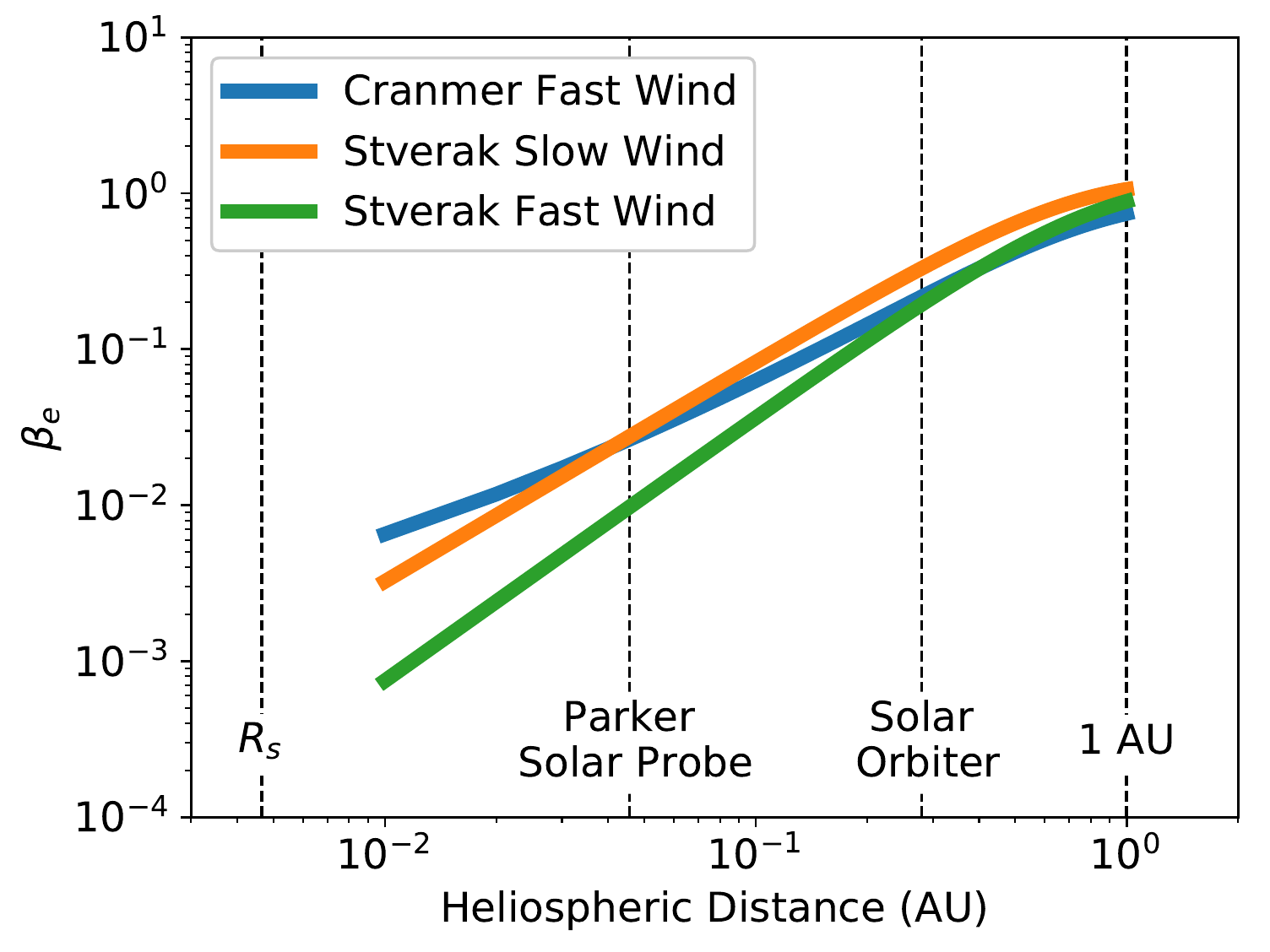}
\caption{{The electron plasma betas in the inner heliosphere. The data are extrapolated back from the existing measurements, using $T_e$ temperature scalings from \citet{cranmer_etal2009} and \citet{stverak2015}, the Parker spiral model for $B$, and the density fits from \citet[][Fig.~1]{bale2016}. The plot indicates a general trend in the change of the parameters, it does not include intrinsic variability of the solar wind parameters and possible changes inside the Alfv\'en radius. The distances accessible by the Solar Probe Plus and Solar Orbiter are shown for reference.}}
\label{Figure1}
\end{figure}

\begin{figure}[h]
\includegraphics[width=\columnwidth]{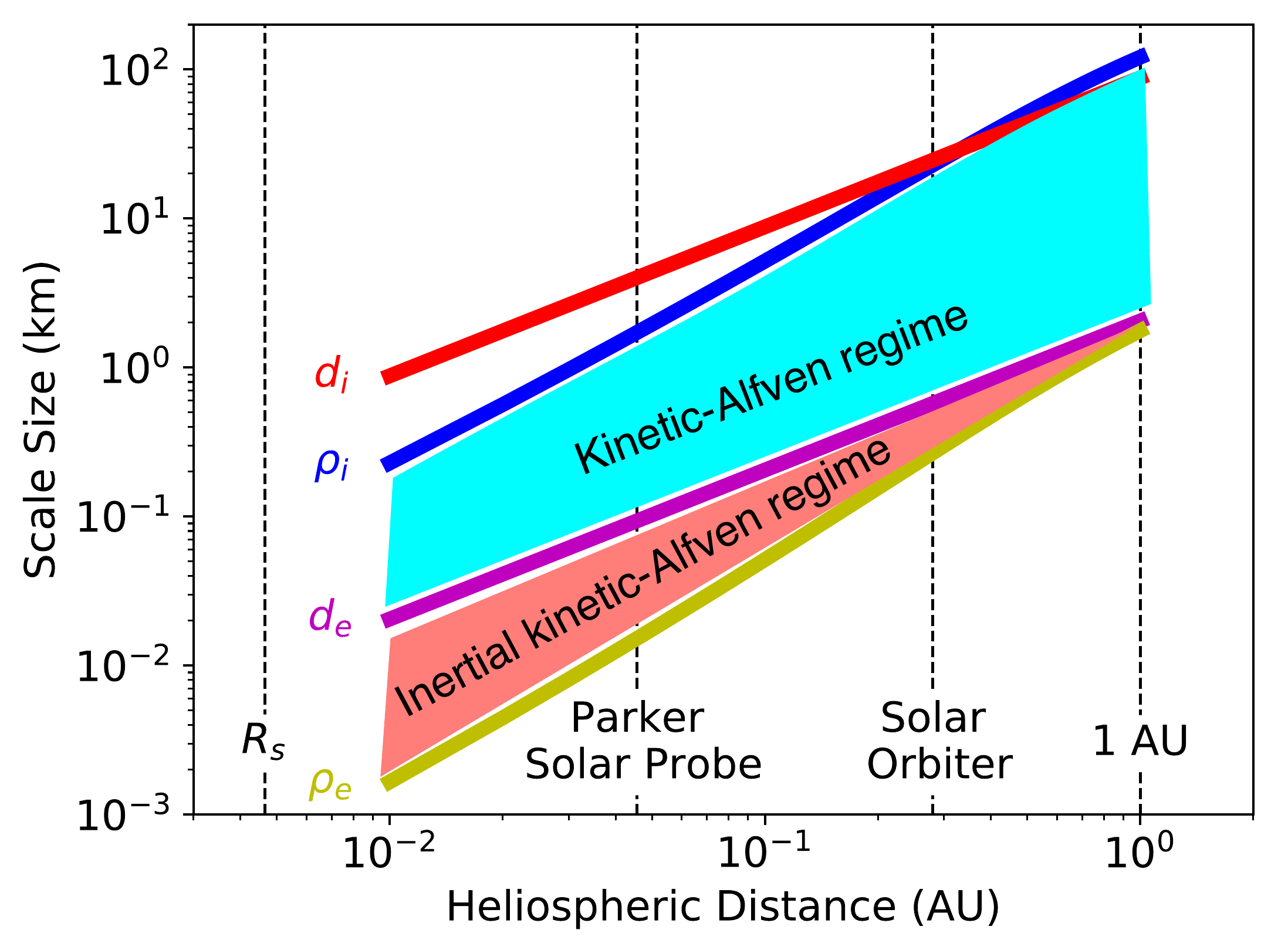}
\caption{{The ion and electron microscales and relevant kinetic plasma modes in the inner heliosphere. The data are extrapolated back from the 1~AU measurements, using the density scaling $\sim 1/r^2$, the Parker spiral model for $B$, and $T_i$, $T_e$ temperature scalings from Helios measurements \cite[][]{cranmer_etal2009}. The plot indicates a general trend in the change of the parameters, it does not include the intrinsic variability of the solar wind parameters and possible changes inside the Alfv\'en radius. The distances accessible by the Solar Probe Plus and Solar Orbiter are shown for reference.}}
\label{Figure2}
\end{figure}

The nature of plasma fluctuations at lower heliospheric distances may however be qualitatively different. 
An important parameter governing their dynamics is the ratio of the plasma thermal energy to the magnetic energy, the plasma beta $\beta_\alpha=8\pi n_\alpha T_\alpha/B^2$ ($\alpha=\{i,e\}$ denotes ions and electrons respectively,  and $n_\alpha$ and $T_\alpha$ are the  particles density and temperature). {Various models and observations-based extrapolations suggest that both the ion and the electron plasma betas decrease in the sunward direction. For example, extrapolations based on temperature scalings in both fast and slow solar wind estimate the electron betas to be about two orders of magnitude smaller than unity in the vicinity of the sun, see Figure~\ref{Figure1}. Similarly, the fast solar wind model of \citet{chandran_etal2011} predicts $\beta_i \sim 0.1$ and $\beta_e \sim 0.01$ at 10 Solar radii from the Sun, while both parameters are of order one ($\beta_i\sim\beta_e\sim 1$) at $1$~AU.} It follows from Fig.~\ref{Figure1} that the plasma beta is decreasing toward the sun, with the electron~$\beta_e$ decreasing faster than the ion~$\beta_i$. Fig.~\ref{Figure2} shows the evolution of {\em ion and electron microscales} in the inner heliosphere estimated based on approximate scalings for plasma parameters. For each particle species, the plasma beta is related to the ratio of their gyroscale to their inertial scale, $\beta_\alpha=\rho^2_\alpha/d^2_\alpha$, {so that the two scales become increasingly well separated when $\beta_\alpha$ decreases.} 

Such missions as Parker Solar Probe and Solar Orbiter will access, for the first time, the lower heliospheric distances, down to  $\sim 9.8$ solar radii. It is therefore highly desirable to develop {theoretical predictions and numerical tools} applicable not only to high heliospheric distances where $\beta \sim 1$, but also to the near-Sun regions where both the ion and electron betas decrease. Besides the low heliospheric distances, such a description will be valuable for other space and astrophysical environments where the electron beta is relatively small, such as the Earth's magnetosheath, interplanetary coronal mass ejections, regions downstream of collisionless shocks, hot accretion flows and others \cite[e.g.,][]{chen_etal2014,treumann09,ghavamian13}.  

{Recently, it has been demonstrated that new low-frequency plasma modes, the so-called inertial kinetic-Alfv\'en (iKAW) modes, appear at the kinetic scales at low values of the electron plasma beta~\cite[][]{chen_boldyrev2017}}. Figures~\ref{Figure1} and~\ref{Figure2} show that the interval of scales {available to these modes} increases progressively as $\beta_e$ decreases.
Analytical models for the {fully nonlinear turbulence in this regime} have been proposed recently by \citet[][]{chen_boldyrev2017} and by~\citet{passot2017}. {In what follows, we will refer to such a regime as the inertial kinetic Alfv\'en turbulence, although we emphasize that this should not be understood to imply that turbulent fluctuations are indeed linear modes. Instead, this change in the nature of fluctuations modifies the kinetic turbulence relative to a better studied regime characterized by $\beta_e \sim 1 $, and may affect the processes of magnetic reconnection, energy dissipation, structure formation, etc. \cite[e.g.,][]{boldyrev_etal2015,chen_boldyrev2017,passot2017,loureiro_boldyrev2017b,mallet2017}. } 

In this paper we present numerical simulations of {plasma turbulence that confirm existence of the iKAW regime}. {We use a fully kinetic formalism that is able to address effects not accounted for by fluid or reduced kinetic models.} Such simulations present formidable challenges to the existing numerical techniques due to the necessity to resolve several disparate micro-scales ($\rho_i$, $d_e$, $\rho_e$) and due to the smallness of the magnetic fluctuations compared to the large-scale magnetic field at $\beta_e\ll 1$. Furthermore, the expectation that the new modes exist both below and above the ion cyclotron frequency~$\Omega_{i}$ necessitates the use of a fully kinetic formalism. In this work, we utilize two complimentary approaches to numerical simulations. First, we discuss a 3D simulation of the dynamics at electron scales using a spectral model based on Fourier-Hermite transform of the full Vlasov-Maxwell system~\citep{Delzanno2015,Vencels2016,Roytershteyn2018}. In order to study the evolution of fluctuations and particle acceleration in a larger system, we also consider a 2D system and conduct a high-resolution Particle-In-Cell (PIC) simulation. {In agreement with the theoretical predictions and recent measurements in the Earth's magnetosheath \cite[][]{chen_boldyrev2017},} we find that the nature of the kinetic turbulence changes from the standard kinetic-Alfv\'en regime at $kd_e<1$ to the new inertial-kinetic-Alfv\'en regime at $kd_e>1$. {In particular, the turbulence spectrum and the magnetic compressibility are different in the new regime and conform to the values predicted by the theory in the appropriate range of scales. The simulations also illuminate certain properties of iKAW turbulence not fully appreciated before, such as the tendency of the fluctuations to develop charge separation at small scales.}  

Throughout the paper, we will use the following notations and definitions: $\omega_{p\alpha} = (4 \pi n_\alpha e^2/m_\alpha)^{1/2}$ and $\Omega_{c\alpha} = e B/(m_\alpha c)$ are plasma and cyclotron frequencies for species $\alpha$ with mass $m_\alpha$ and charge $e$ in magnetic field $B$. The ratio of plasma to magnetic pressure is $\beta_\alpha = 8 \pi n_\alpha T_\alpha/B^2$. The inertial length for species $\alpha$ is denoted by $d_\alpha = c/\omega_{p\alpha}$, their gyroradius is $\rho_\alpha = v_{T\alpha}/\Omega_{c\alpha}$, the thermal speed is $v_{T\alpha} = (2 T_\alpha/m_\alpha)^{1/2}$, and the Debye length is $\lambda_e = T_e^{1/2} /(4 \pi n e^2)^{1/2}$.
 

\section{Inertial Kinetic Alfv\'en Modes}
In this section we summarize the main results of the theory of inertial kinetic-Alfv\'en modes developed by~\cite{chen_boldyrev2017}, which will be used in our study. {This theory is developed in both linear and fully nonlinear regimes.} Consider collisionless plasma in a region permeated by a uniform magnetic field $B_0$ in $z$-direction and  assume that $\beta_e\ll \beta_i\lesssim 1$. Assume that the propagation of waves is oblique, so that the field-parallel and field-perpendicular wave-number components obey $k_\| \ll k_\perp$; such a condition is well satisfied for the small-scale fluctuations in the solar wind \cite[e.g.,][]{mangeney06,alexandrova08c,chen10b,Horbury2012,stawarz16}.  We concentrate on the kinetic scales, i.e. those smaller than the proton gyroscale, $k_\perp^2\rho_i^2\gg 1$. The electromagnetic mode existing in the low-frequency region ($\omega\ll kv_{Ti}$, where $v_{Ti}$ is the proton thermal speed) has been termed the inertial kinetic-Alfv\'en mode by~\cite{chen_boldyrev2017}. Its dispersion relation has the form:
\begin{eqnarray}
\omega^2=\frac{k_\|^2v_\mathrm{A}^2k_\perp^2\rho_\mathrm{i}^2}{\beta_\mathrm{i}(1+k_\perp^2d_\mathrm{e}^2)(1+2/\beta_\mathrm{i}+k_\perp^2d_\mathrm{e}^2)},
\label{eq:ikaw}
\end{eqnarray}  
where $k_\|=k_z$ in the linear case. When the electron inertia effects can be neglected, $k_\perp^2 d_e^2\ll 1$, this mode transforms into a well-studied regular kinetic-Alfv\'en mode \cite[e.g.,][]{howes08a,chen10a,boldyrev12b}. 

The magnetic compressibility for the inertial kinetic Alfv\'en mode is derived as:
\begin{eqnarray}
C_\|^{\rm IKAW} =\frac{\left(\delta B_z\right)^2}{\left(\delta B_x\right)^2+\left(\delta B_y\right)^2}=\frac{1+k_\perp^2d_\mathrm{e}^2}{1+2/\beta_\mathrm{i}+k_\perp^2d_\mathrm{e}^2},
\label{eq:ikawcompressibility}
\end{eqnarray}
where $\delta B_z$ and $\delta B_x, \, \delta B_y$ are field-parallel and field-perpendicular components of the fluctuating  magnetic field.  At $k_\perp^2 d_e^2\ll 1$, the spectrum of strong turbulence dominated by the iKAW modes coincides with the spectrum of the kinetic Alfv\'en modes, which is approximately $E(k)\propto k^{-2.8}$~\cite[e.g.,][]{kiyani09a,chen10b,howes11a,chen12a,boldyrev12b,sahraoui13a,Groselj2018}. It was predicted by \cite{chen_boldyrev2017} that at smaller scales, $k_\perp^2 d_e^2 \gg 1$, the spectrum of magnetic fluctuations steepens to $k^{-11/3}$. {We refer interested readers to~\cite{chen_boldyrev2017} for a detailed discussion of the differences in properties of turbulence between regimes with $\beta \sim 1$ and $\beta \ll 1$.} In this work, we test some of their predictions numerically.

\section{Simulations}
As mentioned in the Introduction, both 2D and 3D numerical approaches are used in our study. The 3D simulations of electron-scale dynamics were conducted using a newly developed version of the spectral Vlasov code SPS~\citep{Delzanno2015,Vencels2016,Roytershteyn2018}. SPS uses an efficient expansion of the distribution function in dual Fourier-Hermite basis and fully implicit time discretization. {The resulting algorithm} possesses exact conservation laws for energy, momentum, and density and is free of unwanted numerical artifacts such as noise. A unique characteristic of the {asymmetrically-weighted} Hermite expansion that the code employs in the velocity space is that {a direct correspondence exists between the evolution equations for the coefficients of expansion and the traditional fluid hierarchy. When a low number of Hermite basis functions is used, the model corresponds to an advanced two-fluid model that is capable of reproducing frequencies and (more qualitatively) damping rates of kinetic Alfv\'en and inertial kinetic Alfv\'en waves~\citep{Roytershteyn2018}. In this work, we consider a 3D simulation performed in a rectangular domain with $L_\parallel = 400\, d_e$ and $L_\perp=50\, d_e$, where $\perp$ and $\parallel$ refer to the direction with respect to the mean field. The simulation is initialized with Maxwellian uniform plasma with density $n_0$. A perturbation of the magnetic field of the form  $\delta {\bm B} = \sum_k \delta {\bm B}_k \cos( \bm k \cdot \bm x + \psi_k)$ and of the ion flow  $\delta {\bm V_i} = \sum_k \delta {\bm V}_k \cos( \bm k \cdot \bm x + \psi_k)$ is imposed at $t=0$, where $\bm k=\{ r (2\pi/L_x), s  (2\pi/L_y), l (2\pi/L_z) \}$, with $r,s = -2\ldots2$ and $l=0\ldots2$. The amplitudes of the individual modes satisfy conditions $\bm k \cdot \delta \bm B_k = 0$, $\bm B_0 \cdot \delta \bm B_k = 0$,  $\bm k \cdot \delta \bm V_k=0$, and $|\delta \bm B_k| = |\delta \bm V_k| $. The mean energy  $\mathcal{E}_0$ of the initial perturbation is $\mathcal{E}_0 = \bar{\mathcal{E}} L_x L_y L_z B_0^2/(8\pi) $, where $\bar{\mathcal{E}} = 0.01$. However, the initial perturbation decays rapidly in time and most of the results discussed below correspond to the times when  $\bar{\mathcal{E}} \sim 0.005$. The background plasma is characterized by $\beta_e = 0.04 $ and $\beta_i \approx 0.4$, so that $T_i/T_e \approx 10$. The ion-to-electron mass ratio is chosen to be $m_i/m_e=100$ and $\omega_{pe}/\Omega_{ce}=2$. The spectral resolution is $N_\perp=255$ in the two perpendicular directions and $N_\parallel=63$ modes in the parallel direction. Four Hermite modes are used in all velocity directions. The time step is $\delta t = 2 \omega_{pe}^{-1}$. The simulations employ an artificial collisional operator to mitigate recurrence issues. The operator used in the present study is defined by Eq.~61 of ~\citep{Delzanno2015} and is constructed to conserve mass, energy, and momentum of each species. The collisionality parameter is $\nu/\omega_{pe}=0.01$. 

Fig.~\ref{fig:sps_linear} demonstrates the ability of the Fourier-Hermite method to capture the behavior of the relevant fluctuations with only 4 Hermite modes per direction in the velocity space, corresponding to a 64-moment system. 
To obtain the frequency and damping rate, we initialized SPS simulations with a low amplitude perturbation corresponding to an eigenvector obtained by numerical solution of the full linearized Vlasov-Maxwell system. The time evolution of the relevant Fourier harmonic of the magnetic field was then fitted to a function of the form $\exp(-\gamma t) \cos( \omega t )$. While the errors in the damping relative to the linear Vlasov result could be significant at high values of $k$ (of the order of 50\%), the overall behavior of the dispersion curves tracks the Vlasov solution very well. {When the number of Hermite basis functions is increased, the damping rates obtained by the Fourier-Hermite method converge to a value that is close to the collisionless damping rate, but is not exactly the same due to the presence of the collision operator. With decreasing value of collisionality, the agreement can be made better at the expense of a 	higher number of Hermite modes needed for convergence. At smaller angles of propagation with respect to the background magnetic field, the measured damping rates are higher than the linear Vlasov results, indicating that oblique modes tend to be significantly overdamped in the simulations with the chosen value of collisionality parameter $\nu$. However, this deviation of the damping rates from the Vlasov predictions does not necessarily represent a serious problem for the simulations reported here since the fluctuations at the electron scales are expected to be nearly perpendicular to the mean magnetic field due to the strong anisotropy of the turbulence cascade.}


\begin{figure}[htbp]
\begin{center}
\includegraphics[width=5in]{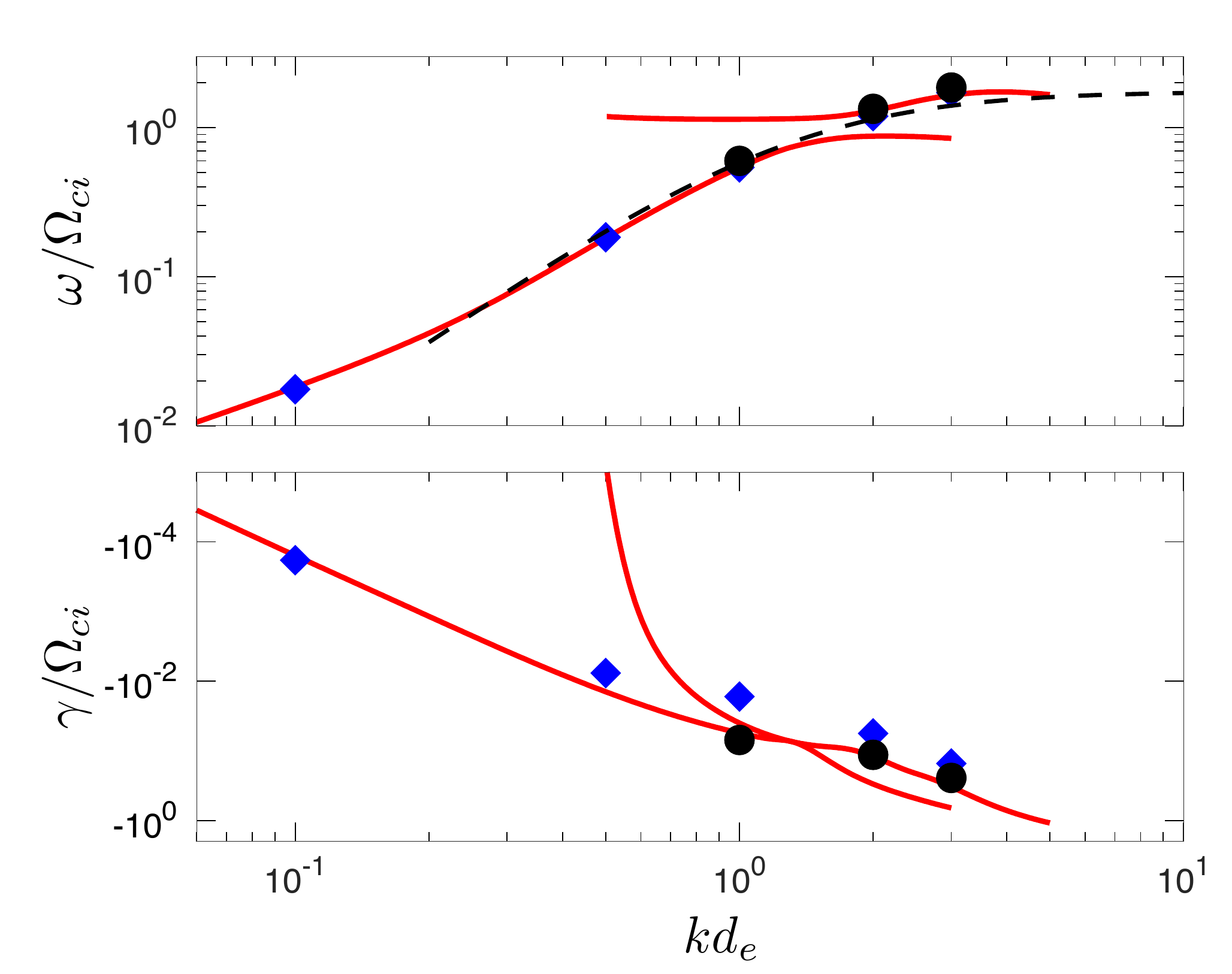}
\caption{An example of the dispersion relation obtained using SPS with a low number of Hermite modes in each direction. Top: frequency, bottom: damping rate. In both panels symbols refer to SPS simulations with 4 (blue diamonds) and 5 (black circles) Hermite modes per direction. Solid curves represent different branches of the dispersion relation obtained with a linear Vlasov solver and the dashed line corresponds to Eq.~\ref{eq:ikaw}. The angle of propagation is $\theta=89^\circ$. Note that similarly to KAW, the iKAW solution crosses over ion Bernstein modes at $\omega > \Omega_{ci}$.}
\label{fig:sps_linear}
\end{center}
\end{figure}

In order to complement the 3D simulations conducted with the truncated Vlasov model, we {also consider} 2D simulations {conducted using} a Particle-In-Cell (PIC) code VPIC~\citep{Bowers2008}. The PIC simulations  include a much wider range of scales and are well-suited for tracking particle energization, which could be significant in low-$\beta$ plasmas.} The simulation plane is perpendicular to the mean magnetic field $B_0$ oriented in the $z$ direction. The plasma parameters are similar to the 3D case: $\beta_e = 0.04$, $\beta_i=0.4$, $\omega_{pe}/\Omega_{ce}=2$, $m_i/m_e=100$. Turbulence is seeded by imposing a randomly-phased perturbation of the type $\delta {\bm B} = \sum_k \delta {\bm B}_k \cos( \bm k \cdot \bm x + \psi_k)$ and $\delta {\bm V} = \sum_k \delta {\bm V}_k \cos( \bm k \cdot \bm x + \phi_k)$ for $\bm k=\{ m (2\pi/L_x),n (2\pi/L_y)\}$, with $m = -2\ldots2$ and $n=0\ldots2$. The size of domain is $L_x = L_y = 8 \pi d_i \approx 251.327\, d_e$ with resolution $n_x = n_y = 3456$ cells, 4000 particles per cell per species, and time step $\omega_{pe}  \delta t \approx 0.05$. 

\section{Results}

\begin{figure}[htbp]
\begin{center}
\includegraphics[width=5in]{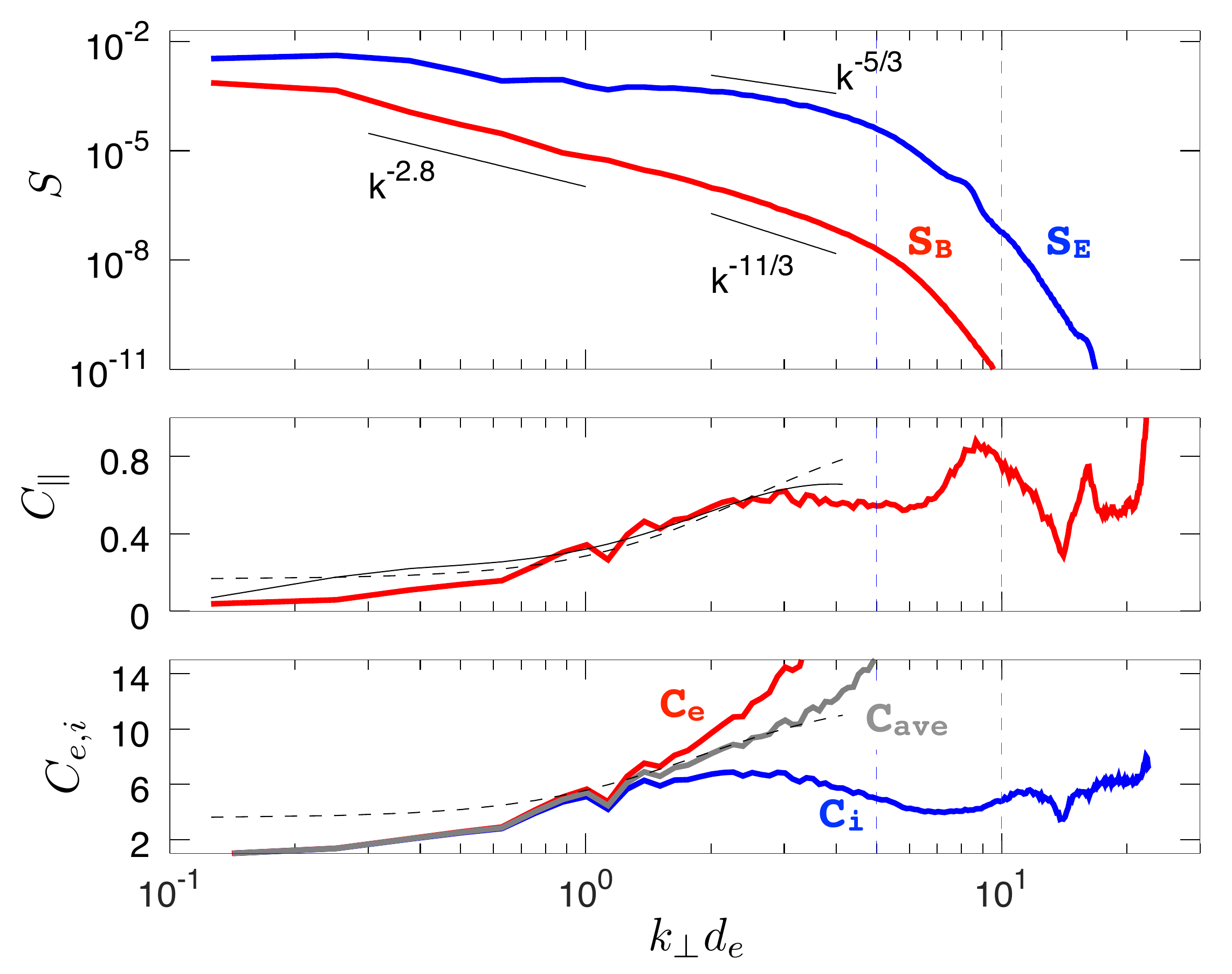}
\caption{{Top: Spectrum of magnetic (B) and electric (E) fluctuations in the 3D SPS simulation at $t\Omega_{ci}=30$, which corresponds approximately to one eddy turnover time. The solid lines show expected slopes $-2.8$ at $k_\perp d_e < 1$,  $-11/3$ at $ d_e^{-1} < k_\perp < \rho_e^{-1}$ for the magnetic fluctuations and $-5/3$ for the electric fluctuations at $ d_e^{-1} < k_\perp < \rho_e^{-1}$.  Middle: parallel compressibility $C_\parallel$. The dashed line shows an analytical prediction for iKAW~\citep{chen_boldyrev2017}, while the thin solid line is a prediction of~\cite{Passot2018} that takes into account electron FLR effects. Bottom: electron ($C_e$) and proton ($C_i$) compressibilities, together with the average $C_{ave} = (C_e+C_i)/2$. The dashed line is the analytical prediction for iKAW. In all panels the vertical lines correspond to scales $k\rho_e=1$ (blue) and $k\lambda_e=1$ (red). The statistical properties shown here remain quasi-stationary during time interval $20 \lesssim t \Omega_{ci} \lesssim 30$.}}
\label{fig:sps_bc}
\end{center}
\end{figure}

The main result of the 3D SPS simulations is summarized in Fig.~\ref{fig:sps_bc}. {The top panel} shows the spectra of magnetic  and electric fluctuations, {together with the typical slopes expected for iKAW turbulence~\cite[][]{chen_boldyrev2017}. In the (short) interval $d_e^{-1} < k_\perp < \rho_e^{-1}$, where the asymptotic theory is applicable, the spectral slopes appear consistent with the theoretical predictions, especially for the magnetic field spectra.} The spectra exhibit sharp steepening at $k_\perp \rho_e \sim 1$, which is consistent with the expected onset of damping at those scales. { The middle panel of Fig.~\ref{fig:sps_bc} shows the values of parallel compressibility $C_\parallel = |\delta B_\parallel|^2/|\delta B_\perp|^2$ measured in the simulations. Here the parallel direction is defined with respect to the mean magnetic field.~\footnote{An alternative definition, which yields almost identical results, is to choose $\delta B_\| = \delta |\bm B|$ as was done by~\cite{chen_boldyrev2017}.} Also shown in the middle panel of Fig.~\ref{fig:sps_bc} are two theoretical predictions, Eq.~\ref{eq:ikawcompressibility} and a prediction of a model that takes into account electron finite Larmor radius (FLR) effects~\cite[Eq.~53 in][]{Passot2018}. Again, the simulation results follow the theoretical predictions well in the range $d_e^{-1} \lesssim k_\perp < \rho_e^{-1}$.  The flattening of the parallel compressibility at $k_\perp \rho_e \approx 0.6$, which for the considered parameters corresponds to  $k_\perp d_e \sim 3$, appears to be an FLR effect. Indeed, the leading FLR correction to Eq.~\ref{eq:ikawcompressibility} introduces a factor $(1-k_\perp^2 \rho_e^2)$ and the solution of~\cite{Passot2018} deviates from Eq.~\ref{eq:ikawcompressibility} approximately at the same value of $k_\perp$. 

The bottom panel in Fig.~\ref{fig:sps_bc} shows electron and ion compressibilities $( \delta n_{e,i}/n_0 )^2 / (|\delta \bm B|/B_0)^2 $. {In contrast to magnetic compressibility, the compressibilities involving density perturbations show differences with the expression given in~\cite{chen_boldyrev2017}. However, the overall trend of compressibility increasing towards small scales, as well as the average value $C_{ave}=(C_e+C_i)/2$ are consistent with the theory.} A likely explanation for this behavior is the charge separation associated with the iKAW fluctuations at $k_\perp d_e \gtrsim 1$, which is exaggerated in simulations where the ratio of the electron plasma frequency to the electron cyclotron frequency is not large.  Indeed, using the solutions in~\cite{chen_boldyrev2017}, it is easy to obtain an estimate $|\delta n_i-\delta n_e|/n_0 \approx (k_\perp d_e)^2 (\Omega_{ce}/\omega_{pe})^2 |\delta B_\parallel|/B_0 $, which is well satisfied in the simulations. 
The degree of charge separation is therefore increased in this simulation relative to majority of examples of low-$\beta$ plasmas cited earlier, since the ratio  $\omega_{pe}/\Omega_{ce}=2$ is significantly lower here than in those systems. However, low values of $\omega_{pe}/\Omega_{ce}=2$ may be encountered in some regions of the solar corona or in the ionosphere~\cite[e.g.][]{Fludra1999,Bilitza2017}. }

\begin{figure}[htbp]
\begin{center}
\includegraphics[width=5in]{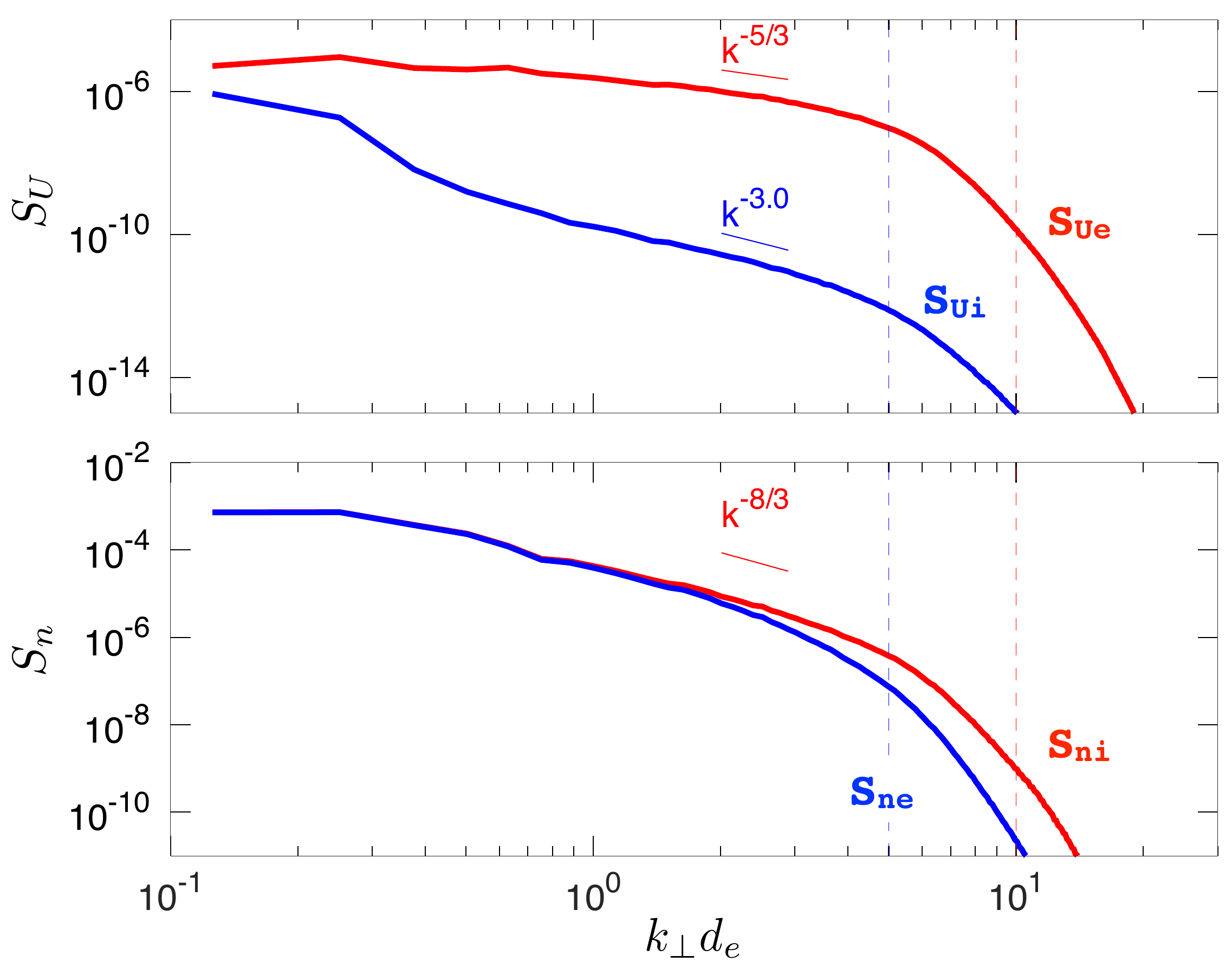}
\caption{Top: spectra of electron (red) and proton (blue) velocity fluctuations. Bottom: spectrum of electron and proton density. The vertical lines denote characteristic scales as in Fig.~\ref{fig:sps_bc}. Various characteristic spectral slopes are shown for reference.}
\label{fig:sps_un}
\end{center}
\end{figure}

The properties of velocity and density fluctuations in the 3D {SPS} simulations are further illustrated by Fig.~\ref{fig:sps_un}. The ion velocity spectrum is much steeper than the electron velocity spectrum, in {qualitative} agreement with previous measurements in the solar wind \cite[e.g.,][]{safrankova13a,safrankova16}, and, in particular, in the Earth's magnetosheath \cite[][Fig. 4]{chen_boldyrev2017}. The spectrum of electron density fluctuations, shown in the bottom panel of Fig.~\ref{fig:sps_un}, is significantly shallower than the spectrum of the ion fluctuations, leading to the quasi-neutrality violations discussed above. 

\begin{figure}[htbp]
\begin{center}
\includegraphics[width=5in]{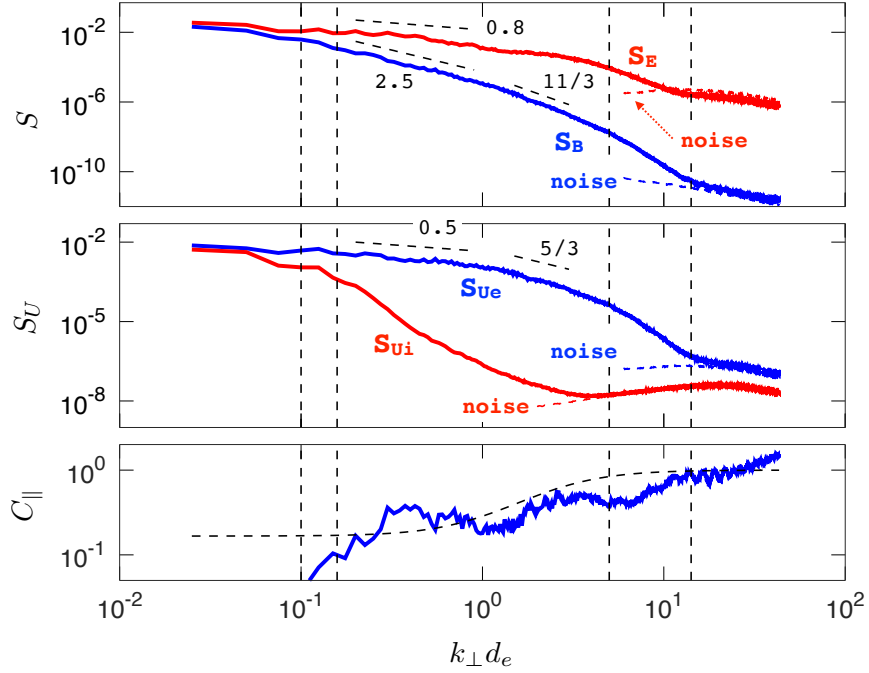}
\caption{Top panel: spectrum of magnetic (blue) and electric (red) fluctuations in the 2D PIC simulation at $t\Omega_{ci}\approx126$, when the turbulence is developed. Middle panel: spectra of ion (red) and electron (blue) velocity fluctuations.  Bottom panel: magnetic compressibility $C_\parallel = |\delta B_z|^2/(|\delta B_x|^2+|\delta B_y|^2)$ at $t\Omega_{ci}\approx126$. The dashed line shows the analytic prediction~(\ref{eq:ikawcompressibility}). In all panels the vertical lines mark scales corresponding to (in order of increasing $k_\perp$) $k_\perp{d_i}=1$, $k_\perp\rho_i=1$, $k_\perp\rho_e = 1$, and $k_\perp\lambda_e = 1$. Spectra are computed from fields averaged over 50 steps, corresponding to time interval $T_{av}\Omega_{ce} \approx 1.26$. The top and middle panels also include various characteristic power laws $S\sim k^{-\zeta}$, labeled by the value of the index $\zeta$.}
\label{fig:kspectrum}
\end{center}
\end{figure}

In order to investigate the behavior of low-$\beta$ turbulence in a much larger system and with larger resolution than is currently feasible in 3D {SPS} simulations, we turn attention to the 2D PIC simulations. The top panel in Fig.~\ref{fig:kspectrum} shows the spectrum of magnetic and electric fluctuations at the time $t\Omega_{ci}=125.7$, approximately corresponding to one eddy turnover time.  By this time, approximately 40\% of the initial perturbation energy has decayed and the statistical properties of fluctuations are approximately stationary. As expected, we can identify three characteristic ranges of scales in the energy spectrum. At scales $k_\perp d_e \lesssim 1$, the magnetic spectra are relatively flat, with the {local slopes} approximately equal to~-2.5, which is close to the power law indices~-2.8 frequently observed in data and simulations of kinetic-Alfv\'en turbulence \cite[e.g.,][]{alexandrova09,kiyani09a,chen10b,howes11a,chen12a,boldyrev12b,sahraoui13a,Groselj2018}. After $kd_e \sim 1$, the spectra gradually steepen, with local slopes approaching $3.5-4.0$ in the range of scales $ d_e^{-1} \lesssim k < \rho_e^{-1}$. This behavior is consistent with the prediction for the inertial kinetic Alfv\'en turbulence, with 3D {SPS} simulations presented earlier, and with the recent observations of turbulence in the Earth's magnetosheath \cite[][Fig. 5]{chen_boldyrev2017}.  Finally, another transition is observed at $k_\perp \rho_e \sim 1 $, where the spectral index further increases to approximately~6.5. 

The spectra of ion and electron flows, shown in the middle panel of Fig.~\ref{fig:kspectrum}, are consistent with the results of 3D simulations. {In particular, the ion flow fluctuations have substantially lower amplitude than the electron ones and the spectral indices for the electron flow correspond to those of the current density}. The magnetic compressibility measured in simulations is plotted in the bottom panel of Fig.~\ref{fig:kspectrum}, together with the predicted curve for the iKAW given by Eq.~(\ref{eq:ikawcompressibility}). We see that compressibility {exhibits the same trend as} {shown by the 3D simulations described earlier}.  At the same time, the compressibility does not follow the theoretical prediction as cleanly in this 2D PIC simulation as it does in the 3D SPS case.
 
\begin{figure}[htbp]
\begin{center}
\includegraphics[width=3in]{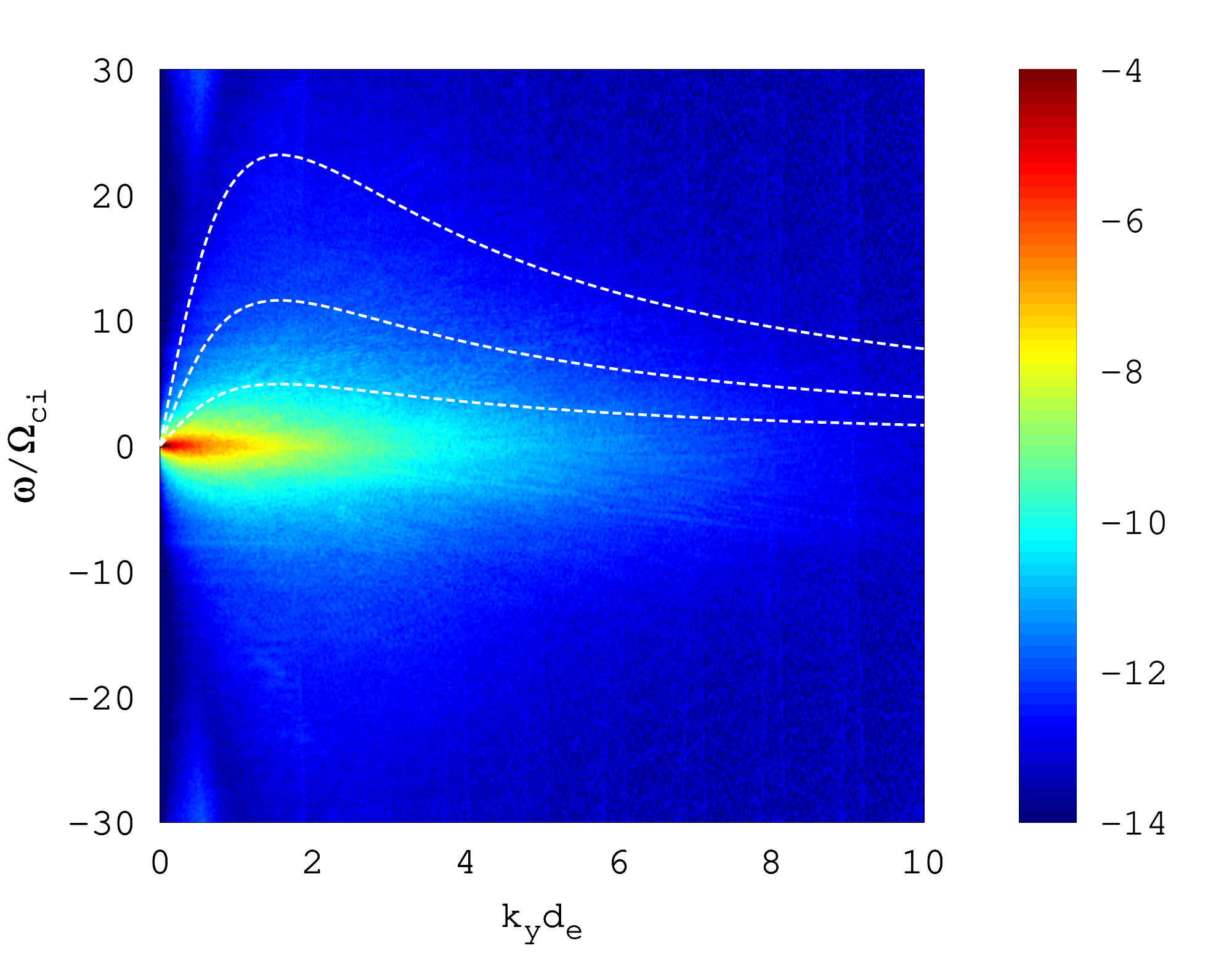}
\caption{Frequency spectrum of magnetic fluctuations $\log_{10} |B_x(\omega,k_y)|^2$ in a 2D PIC simulation. The dashed lines show dispersion relation for iKAW modes given by Eq.~(\ref{eq:ikaw}) as a function of $k_\perp$ for several values of $k_z$.}
\label{fig:wkspectrum}
\end{center}
\end{figure}

Further evidence that the small-scale fluctuations in the 2D simulation exhibit some properties expected of inertial kinetic-Alfv\'en turbulence is supplied by analysis of the frequency spectrum of magnetic fluctuations. Fig.~\ref{fig:wkspectrum} shows an example of such a spectrum obtained  by a number of high-frequency ``probes" in the simulation. The probes are arranged in a $16 \times 3456$ grid and record the values of all 6 components of electromagnetic field. The maximum frequency resolved by this diagnostic in the simulation presented here corresponds to the electron cyclotron frequency. The spectra shown in Fig.~\ref{fig:wkspectrum} were obtained in the interval $ 100 \lesssim t \Omega_{ci} \lesssim  150$ by taking a 2D Fourier transform in time and in the $y$ direction and averaging over 16 locations in $x$. Overplotted on the spectrum are several curves corresponding to the dispersion relation for iKAW modes (\ref{eq:ikaw}) at fixed $k_\|$. The {shapes of constant level regions for frequencies both below and above $\Omega_{ci}$ (but well below $\Omega_{ce}$) are well described} by the analytical expression. We believe that the large-scale perturbations of the magnetic field enable finite values of $k_\|$ to exist {\em locally}, which explains existence of the iKAW fluctuations at small scales in this 2D simulation with an out-of-plane mean magnetic field. 

Parenthetically, we note that in addition to the fluctuations demonstrated in Fig.~\ref{fig:wkspectrum},  we have also detected in our simulations high frequency magnetic and electric fluctuations of significant amplitude in two frequency bands, below and above the electron plasma frequency at spatial scales $kd_e \lesssim 1$ (not shown here). They are consistent with the $X$-modes and electron Bernstein mode continuum.  These fluctuations are however not a part of kinetic Alfv\'en turbulence. They appear to be excited by the decay of the initial perturbation, which was chosen to be rather general, not merely consisting of kinetic Alfv\'en eigenmodes. The high-frequency fluctuations were generated at relatively early times $t\Omega_{ci} \lesssim 20$, and persisted over the duration of simulation with moderate reduction in amplitude. 
 
One of the interesting questions {regarding transition to iKAW turbulence} is what consequences, if any, the transition has on the mechanisms of turbulent energy dissipation. While a full investigation of these issues is beyond the scope of this paper, we briefly discuss the properties of current structures observed in the 2D simulation. 

\begin{figure*}[htbp]
\begin{center}
\includegraphics[width=\textwidth]{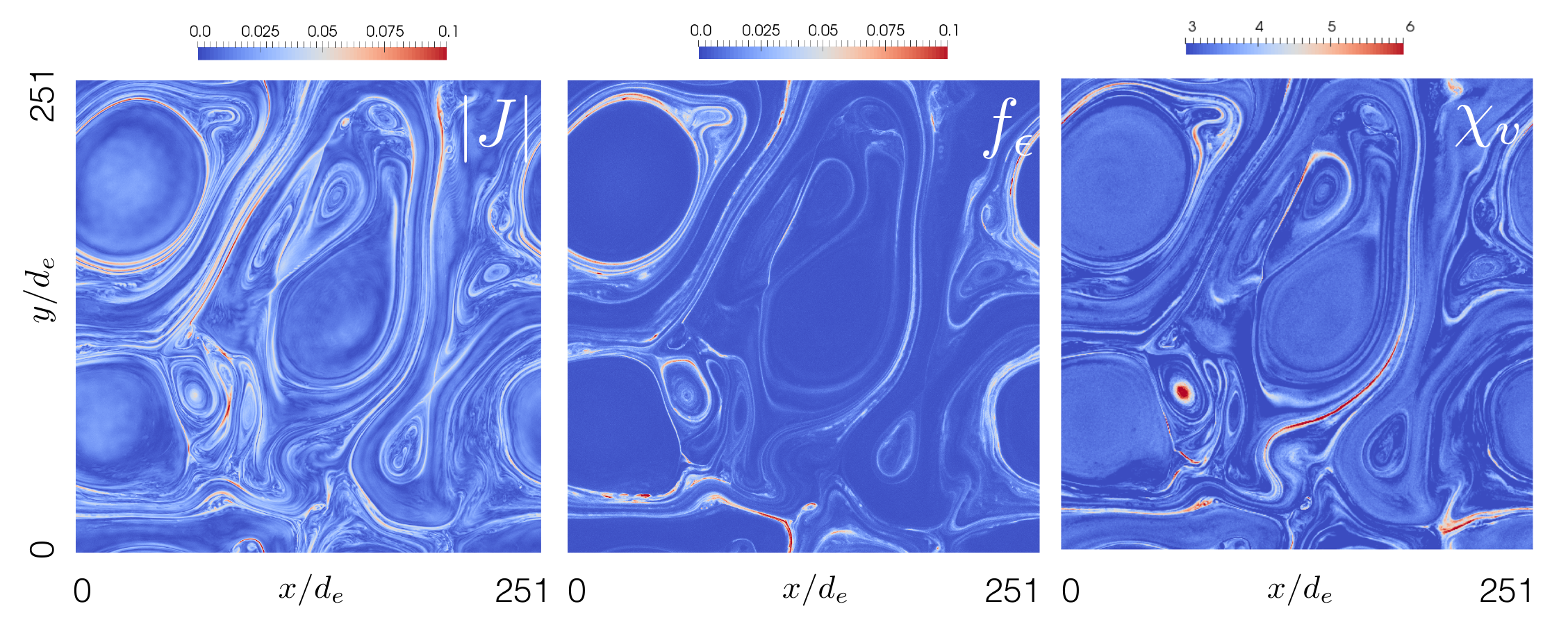}
\caption{Left: current density in 2D PIC simulation at time $t\Omega_{ci}\approx126$. Middle: fraction of local density taken by energetic electrons with energy $\epsilon \geq 8 T_e^0$. Right: local kurtosis of electron distribution functions. Values greater than 3 indicate deviation of the electron distribution from Maxwellian. See text for more details.}
\label{fig:structures}
\end{center}
\end{figure*}

{As is well known, plasma turbulence tends to generate strong current layers~\citep[e.g.][]{Matthaeus1986,Biskamp1989}. In kinetic plasmas, the current sheets span a range of scales down to $d_e$~\citep[e.g.][]{karimabadi2013}. Moreover, reconnecting current sheets can develop sub-structure on scales comparable to the electron gyroradius~\cite[e.g.][]{Ricci2004,Liu2013}. Kinetic simulations suggest that the current layers can play a significant role in the overall energy dissipation~\cite[e.g.][]{Wan2012,karimabadi2013,Tenbarge2013,Wan2016,Camporeale2018}. Figure~\ref{fig:structures} illustrates properties of current structures in the 2D PIC simulation.  Such current sheets are often unstable to tearing and other micro-instabilities, which may modify turbulent dynamics~\citep[e.g.][]{Carbone1990,Mallet2016,Loureiro2017,loureiro_boldyrev2017b}. Further, in a weakly collisional plasma, the development of thin current sheets and onset of reconnection may lead to large deviation from a Maxwellian in the particle distribution functions.} The middle panel in Fig.~\ref{fig:structures} illustrates local density fraction of high-energy particles, $f_\epsilon = \int_\epsilon^{\infty} F(\epsilon)\,d\epsilon / \int_{-\infty}^{\infty} F(\epsilon)\,d\epsilon$, where $F(\epsilon)$ is the distribution function in energy $\epsilon$. Evidently, the high-energy particles tend to be generated and accumulated in the vicinity of narrow current structures. The right panel shows local kurtosis of the particle distribution function $F(v)$, where values exceeding 3 indicate appearance of non-Maxwellian distributions. The non-Maxwellian features are also strongly localized in relatively narrow regions.

\section{Discussion and Conclusions}
Recent observational and analytical studies point to a new type of magnetic plasma turbulence existing in the regime of low electron beta $\beta_e\ll \beta_i\lesssim 1$, the so-called inertial kinetic Alfv\'en turbulence \cite[][]{chen_boldyrev2017,passot2017,Passot2018}. {The predicted properties of such turbulence deviate significantly from the properties of the much better studied {regime characterized by no separation of scales between electron inertial length and the electron gyroradius (i.e. $\beta_e \sim 1$), which is often referred to as }kinetic Alfv\'en turbulence. For example, iKAW turbulence is expected to be characterized by different spectra of magnetic fluctuations as well as different degree of anisotropy. Furthermore, the frequency of iKAW fluctuations  may significantly exceed the ion cyclotron frequency. All of these properties may change mechanisms of energy dissipation, processes of structure formation, particle acceleration, and magnetic reconnection relative to KAW turbulence. {We emphasize that the regimes of interest here correspond to strong turbulence, i.e. there is no expectation that the turbulent fluctuations are linear plasma modes even if such fluctuations are expected to retain some characteristics of linear modes. Examples of such correspondence between linear and nonlinear phenomena are abundant in plasma physics in general and in plasma turbulence studies in particular~\citep[e.g.][]{Matthaeus1991,Goldreich1995}. Furthermore, there exists ample observational evidence in the solar wind suggesting that in a statistical sense turbulent fluctuations retain some characteristics of the corresponding linear modes at both large magnetohydrodynamic and small kinetic scales~\cite[e.g.][]{Belcher1971,Salem2012,Chen2013}.}

In this work we have presented the first numerical simulations of iKAW turbulence. The simulations utilized a fully kinetic formalism and as such are well suited for studying the transition to iKAW turbulence as well as the properties of such turbulence. As a first step towards better understanding of iKAW turbulence, our initial simulations focused predominantly on verifying key predictions of the theory.  The simulations confirmed the existence of a transition at scales corresponding to $k_\perp d_e \sim 1 $. While the range of scales where iKAW turbulence can develop $d_e^{-1} < k_\perp < \rho_e^{-1}$ is limited in our simulations due to only moderately low value of $\beta_e =0.04$ considered, in that range the simulation results are consistent with the analytic predictions for the value of spectral slope of magnetic fluctuations, as well as the value of magnetic compressibility. Similar agreement was found recently between in situ plasma measurements in the Earth's magnetosheath and the theory. The results presented here lend support to the conclusion made by \cite{chen_boldyrev2017} that they observed a new type of low-frequency plasma turbulence. The simulations also revealed certain effects not appreciated earlier. For example, the behavior of electron and ion compressibilities is affected by the tendency of iKAW fluctuations to develop charge separation at small scales. This charge separation is exaggerated in the simulations due to the utilized value of the parameter $\omega_{pe}/\omega_{ce} = 2$ and should be much smaller in the magnetosheath and other systems where $\omega_{pe}/\omega_{ce}$ is significantly larger. At the same time, the charge separation may be significant in the ionosphere or in coronal holes. 

{We emphasize that the increased separation of scales typical of the regimes with low value of $\beta$ makes kinetic simulations extremely challenging. The results presented here are only the first steps towards  a more comprehensive understanding of the turbulence in such regimes. The presented simulations focus on a relatively small range of scales near the electron kinetic scales, thus excluding many aspects of the turbulent dynamics that are associated with large-scale motion. Further progress requires continuing development of advanced algorithms capable of efficiently simulating multiscale dynamics in low-$\beta$ plasmas.}




\section{Acknowledgments}
VR was partly supported by NASA grant NNX15AR16G. SB was partly supported by the NSF grant No. PHY-1707272, NASA grant No. 80NSSC18K0646 and by the Vilas Associates Award from the University of Wisconsin - Madison. GLD acknowledges funding by the Laboratory Directed Research and Development (LDRD) program, under the auspices of the National Nuclear Security Administration of the U.S. Department of Energy by Los Alamos National Laboratory, operated by Los Alamos National Security LLC under contract DE-AC52-06NA25396. CHKC is supported by STFC Ernest Rutherford Fellowship ST/N003748/2. NFL was funded by the NSF CAREER award no. 1654168

Computational resources were provided by the NASA High-End Computing Program through the NASA Advanced Supercomputing Division at Ames Research Center. PIC simulations were conducted as a part of the Blue Waters sustained-petascale computing project, which is supported by the National Science Foundation (awards OCI-0725070 and ACI-1238993) and the state of Illinois. Blue Waters is a joint effort of the University of Illinois at Urbana-Champaign and its National Center for Supercomputing Applications. Blue Waters allocation was provided by the National Science Foundation through PRAC award 1614664.


\bibliography{low_beta_paper}

\begin{thebibliography}{}
\expandafter\ifx\csname natexlab\endcsname\relax\def\natexlab#1{#1}\fi
\providecommand{\url}[1]{\href{#1}{#1}}
\providecommand{\dodoi}[1]{doi:~\href{http://doi.org/#1}{\nolinkurl{#1}}}
\providecommand{\doeprint}[1]{\href{http://ascl.net/#1}{\nolinkurl{http://ascl.net/#1}}}
\providecommand{\doarXiv}[1]{\href{https://arxiv.org/abs/#1}{\nolinkurl{https://arxiv.org/abs/#1}}}

\bibitem[{{Alexandrova} {et~al.}(2008){Alexandrova}, {Lacombe}, \&
  {Mangeney}}]{alexandrova08c}
{Alexandrova}, O., {Lacombe}, C., \& {Mangeney}, A. 2008, \ang, 26, 3585,
  \dodoi{10.5194/angeo-26-3585-2008}

\bibitem[{{Alexandrova} {et~al.}(2009){Alexandrova}, {Saur}, {Lacombe},
  {Mangeney}, {Mitchell}, {Schwartz}, \& {Robert}}]{alexandrova09}
{Alexandrova}, O., {Saur}, J., {Lacombe}, C., {et~al.} 2009, Physical Review
  Letters, 103, 165003, \dodoi{10.1103/PhysRevLett.103.165003}

\bibitem[{{Bale} {et~al.}(2016){Bale}, {Goetz}, {Harvey}, {Turin}, {Bonnell},
  {Dudok de Wit}, {Ergun}, {MacDowall}, {Pulupa}, {Andre}, {Bolton},
  {Bougeret}, {Bowen}, {Burgess}, {Cattell}, {Chandran}, {Chaston}, {Chen},
  {Choi}, {Connerney}, {Cranmer}, {Diaz-Aguado}, {Donakowski}, {Drake},
  {Farrell}, {Fergeau}, {Fermin}, {Fischer}, {Fox}, {Glaser}, {Goldstein},
  {Gordon}, {Hanson}, {Harris}, {Hayes}, {Hinze}, {Hollweg}, {Horbury},
  {Howard}, {Hoxie}, {Jannet}, {Karlsson}, {Kasper}, {Kellogg}, {Kien},
  {Klimchuk}, {Krasnoselskikh}, {Krucker}, {Lynch}, {Maksimovic}, {Malaspina},
  {Marker}, {Martin}, {Martinez-Oliveros}, {McCauley}, {McComas}, {McDonald},
  {Meyer-Vernet}, {Moncuquet}, {Monson}, {Mozer}, {Murphy}, {Odom},
  {Oliverson}, {Olson}, {Parker}, {Pankow}, {Phan}, {Quataert}, {Quinn},
  {Ruplin}, {Salem}, {Seitz}, {Sheppard}, {Siy}, {Stevens}, {Summers}, {Szabo},
  {Timofeeva}, {Vaivads}, {Velli}, {Yehle}, {Werthimer}, \&
  {Wygant}}]{bale2016}
{Bale}, S.~D., {Goetz}, K., {Harvey}, P.~R., {et~al.} 2016, Space Science
  Reviews, 204, 49, \dodoi{10.1007/s11214-016-0244-5}

\bibitem[{Belcher \& {Davis, Jr.}(1971)}]{Belcher1971}
Belcher, J.~W., \& {Davis, Jr.}, L. 1971, Journal of Geophysical Research, 76,
  3534

\bibitem[{Bilitza {et~al.}(2017)Bilitza, Altadill, Truhlik, Shubin, Galkin,
  Reinisch, \& Huang}]{Bilitza2017}
Bilitza, D., Altadill, D., Truhlik, V., {et~al.} 2017, Space Weather, 15, 418,
  \dodoi{10.1002/2016SW001593}

\bibitem[{Biskamp \& Welter(1989)}]{Biskamp1989}
Biskamp, D., \& Welter, H. 1989, Physics of Fluids B, 1, 1964,
  \dodoi{10.1063/1.859060}

\bibitem[{Boldyrev {et~al.}(2015)Boldyrev, Chen, Xia, \&
  Zhdankin}]{boldyrev_etal2015}
Boldyrev, S., Chen, C. H.~K., Xia, Q., \& Zhdankin, V. 2015, The Astrophysical
  Journal, 806, 238, \dodoi{10.1088/0004-637X/806/2/238}

\bibitem[{{Boldyrev} \& {Perez}(2012)}]{boldyrev12b}
{Boldyrev}, S., \& {Perez}, J.~C. 2012, The Astrophysical Journall, 758, L44,
  \dodoi{10.1088/2041-8205/758/2/L44}

\bibitem[{Bowers {et~al.}(2008)Bowers, Albright, Yin, Bergen, \&
  Kwan}]{Bowers2008}
Bowers, K.~J., Albright, B.~J., Yin, L., Bergen, B., \& Kwan, T. J.~T. 2008,
  Physics of Plasmas, 15, 055703, \dodoi{10.1063/1.2840133}

\bibitem[{Bruno \& Carbone(2005)}]{bruno:2005}
Bruno, R., \& Carbone, V. 2005, Living Reviews in Solar Physics, 2,
  \dodoi{10.1007/lrsp-2005-4}

\bibitem[{Camporeale {et~al.}(2018)Camporeale, Sorriso-Valvo, Califano, \&
  Retin\`o}]{Camporeale2018}
Camporeale, E., Sorriso-Valvo, L., Califano, F., \& Retin\`o, A. 2018, Physical
  Review Letters, 120, 125101, \dodoi{10.1103/PhysRevLett.120.125101}

\bibitem[{Carbone {et~al.}(1990)Carbone, Veltri, \& Mangeney}]{Carbone1990}
Carbone, V., Veltri, P., \& Mangeney, A. 1990, Physics of Fluids A, 2, 1487,
  \dodoi{10.1063/1.857598}

\bibitem[{Chandran {et~al.}(2011)Chandran, Dennis, Quataert, \&
  Bale}]{chandran_etal2011}
Chandran, B. D.~G., Dennis, T.~J., Quataert, E., \& Bale, S.~D. 2011, The
  Astrophysical Journal, 743, 197, \dodoi{10.1088/0004-637X/743/2/197}

\bibitem[{Chen {et~al.}(2013)Chen, Boldyrev, Xia, \& Perez}]{Chen2013}
Chen, C.~H., Boldyrev, S., Xia, Q., \& Perez, J.~C. 2013, Physical Review
  Letters, 110, 1, \dodoi{10.1103/PhysRevLett.110.225002}

\bibitem[{Chen {et~al.}(2014)Chen, Leung, Boldyrev, Maruca, \&
  Bale}]{chen_etal2014}
Chen, C.~H., Leung, L., Boldyrev, S., Maruca, B.~A., \& Bale, S.~D. 2014,
  Geophysal Research Letters, 41, 8081, \dodoi{10.1002/2014GL062009}

\bibitem[{Chen \& Boldyrev(2017)}]{chen_boldyrev2017}
Chen, C. H.~K., \& Boldyrev, S. 2017, The Astrophysical Journal, 842, 122,
  \dodoi{10.3847/1538-4357/aa74e0}

\bibitem[{{Chen} {et~al.}(2010{\natexlab{a}}){Chen}, {Horbury}, {Schekochihin},
  {Wicks}, {Alexandrova}, \& {Mitchell}}]{chen10b}
{Chen}, C.~H.~K., {Horbury}, T.~S., {Schekochihin}, A.~A., {et~al.}
  2010{\natexlab{a}}, Physical Review Letters, 104, 255002,
  \dodoi{10.1103/PhysRevLett.104.255002}

\bibitem[{{Chen} {et~al.}(2012){Chen}, {Salem}, {Bonnell}, {Mozer}, \&
  {Bale}}]{chen12a}
{Chen}, C.~H.~K., {Salem}, C.~S., {Bonnell}, J.~W., {Mozer}, F.~S., \& {Bale},
  S.~D. 2012, Physical Review Letters, 109, 035001,
  \dodoi{10.1103/PhysRevLett.109.035001}

\bibitem[{{Chen} {et~al.}(2010{\natexlab{b}}){Chen}, {Wicks}, {Horbury}, \&
  {Schekochihin}}]{chen10a}
{Chen}, C.~H.~K., {Wicks}, R.~T., {Horbury}, T.~S., \& {Schekochihin}, A.~A.
  2010{\natexlab{b}}, The Astrophysical Journall, 711, L79,
  \dodoi{10.1088/2041-8205/711/2/L79}

\bibitem[{Cranmer {et~al.}(2009)Cranmer, Matthaeus, Breech, \&
  Kasper}]{cranmer_etal2009}
Cranmer, S., Matthaeus, W., Breech, B., \& Kasper, J. 2009, The Astrophysical
  Journal, 702, 1604, \dodoi{10.1088/0004-637X/702/2/1604}

\bibitem[{Delzanno(2015)}]{Delzanno2015}
Delzanno, G. 2015, Journal of Computational Physics, 301, 338,
  \dodoi{10.1016/j.jcp.2015.07.028}

\bibitem[{Fludra {et~al.}(1999)Fludra, {Del Zanna}, Alexander, \&
  Bromage}]{Fludra1999}
Fludra, A., {Del Zanna}, G., Alexander, D., \& Bromage, B. J.~I. 1999, Journal
  of Geophysical Research: Space Physics, 104, 9709,
  \dodoi{10.1029/1998JA900033}

\bibitem[{{Ghavamian} {et~al.}(2013){Ghavamian}, {Schwartz}, {Mitchell},
  {Masters}, \& {Laming}}]{ghavamian13}
{Ghavamian}, P., {Schwartz}, S.~J., {Mitchell}, J., {Masters}, A., \& {Laming},
  J.~M. 2013, Space Science Reviews, 178, 633,
  \dodoi{10.1007/s11214-013-9999-0}

\bibitem[{Goldreich \& Sridhar(1995)}]{Goldreich1995}
Goldreich, P., \& Sridhar, S. 1995, The Astrophysical Journal, 438, 763,
  \dodoi{10.1086/175121}

\bibitem[{Gro{\v{s}}elj {et~al.}(2018)Gro{\v{s}}elj, Mallet, Loureiro, \&
  Jenko}]{Groselj2018}
Gro{\v{s}}elj, D., Mallet, A., Loureiro, N.~F., \& Jenko, F. 2018, Physical
  Review Letters, 120, 1, \dodoi{10.1103/PhysRevLett.120.105101}

\bibitem[{Horbury {et~al.}(2012)Horbury, Wicks, \& Chen}]{Horbury2012}
Horbury, T.~S., Wicks, R.~T., \& Chen, C. H.~K. 2012, Space Science Reviews,
  172, 325, \dodoi{10.1007/s11214-011-9821-9}

\bibitem[{{Howes} {et~al.}(2008){Howes}, {Cowley}, {Dorland}, {Hammett},
  {Quataert}, \& {Schekochihin}}]{howes08a}
{Howes}, G.~G., {Cowley}, S.~C., {Dorland}, W., {et~al.} 2008, Journal of
  Geophysical Research, 113, 5103, \dodoi{10.1029/2007JA012665}

\bibitem[{{Howes} {et~al.}(2011){Howes}, {TenBarge}, {Dorland}, {Quataert},
  {Schekochihin}, {Numata}, \& {Tatsuno}}]{howes11a}
{Howes}, G.~G., {TenBarge}, J.~M., {Dorland}, W., {et~al.} 2011, Physical
  Review Letters, 107, 035004, \dodoi{10.1103/PhysRevLett.107.035004}

\bibitem[{Karimabadi {et~al.}(2013)Karimabadi, Roytershteyn, Wan, Matthaeus,
  Daughton, Wu, Shay, Loring, Borovsky, Leonardis, Chapman, \&
  Nakamura}]{karimabadi2013}
Karimabadi, H., Roytershteyn, V., Wan, M., {et~al.} 2013, Physics of Plasmas,
  20, 12303, \dodoi{10.1063/1.4773205}

\bibitem[{{Kiyani} {et~al.}(2009){Kiyani}, {Chapman}, {Khotyaintsev}, {Dunlop},
  \& {Sahraoui}}]{kiyani09a}
{Kiyani}, K.~H., {Chapman}, S.~C., {Khotyaintsev}, Y.~V., {Dunlop}, M.~W., \&
  {Sahraoui}, F. 2009, Physical Review Letters, 103, 075006,
  \dodoi{10.1103/PhysRevLett.103.075006}

\bibitem[{Kiyani {et~al.}(2015)Kiyani, Osman, \& Chapman}]{Kiyani2015}
Kiyani, K.~H., Osman, K.~T., \& Chapman, S.~C. 2015, Philosophical Transactions
  of the Royal Society A: Mathematical, Physical and Engineering Sciences, 373,
  20140155, \dodoi{10.1098/rsta.2014.0155}

\bibitem[{Liu {et~al.}(2013)Liu, Daughton, Karimabadi, Li, \&
  Roytershteyn}]{Liu2013}
Liu, Y.-H., Daughton, W., Karimabadi, H., Li, H., \& Roytershteyn, V. 2013,
  Physical Review Letters, 110, 265004, \dodoi{10.1103/PhysRevLett.110.265004}

\bibitem[{Loureiro \& Boldyrev(2017{\natexlab{a}})}]{loureiro_boldyrev2017b}
Loureiro, N.~F., \& Boldyrev, S. 2017{\natexlab{a}}, The Astrophysical Journal,
  850, 182, \dodoi{10.3847/1538-4357/aa9754}

\bibitem[{Loureiro \& Boldyrev(2017{\natexlab{b}})}]{Loureiro2017}
---. 2017{\natexlab{b}}, Physical Review Letters, 118, 1,
  \dodoi{10.1103/PhysRevLett.118.245101}

\bibitem[{Mallet {et~al.}(2017{\natexlab{a}})Mallet, Schekochihin, \&
  Chandran}]{mallet2017}
Mallet, A., Schekochihin, A.~A., \& Chandran, B. D.~G. 2017{\natexlab{a}},
  Journal of Plasma Physics, 83, 905830609, \dodoi{10.1017/S0022377817000812}

\bibitem[{Mallet {et~al.}(2017{\natexlab{b}})Mallet, Schekochihin, \&
  Chandran}]{Mallet2016}
---. 2017{\natexlab{b}}, Monthly Notices of the Royal Astronomical Society,
  468, 4862, \dodoi{10.1093/mnras/stx670}

\bibitem[{{Mangeney} {et~al.}(2006){Mangeney}, {Lacombe}, {Maksimovic},
  {Samsonov}, {Cornilleau-Wehrlin}, {Harvey}, {Bosqued}, \&
  {Tr{\'a}vn{\'{\i}}{\v c}ek}}]{mangeney06}
{Mangeney}, A., {Lacombe}, C., {Maksimovic}, M., {et~al.} 2006, \ang, 24, 3507,
  \dodoi{10.5194/angeo-24-3507-2006}

\bibitem[{Matthaeus {et~al.}(1991)Matthaeus, Klein, Ghosh, \&
  Brown}]{Matthaeus1991}
Matthaeus, W., Klein, L.~W., Ghosh, S., \& Brown, M.~R. 1991, Journal of
  Geophysical Research, 96, 5421, \dodoi{10.1029/90JA02609}

\bibitem[{Matthaeus \& Lamkin(1986)}]{Matthaeus1986}
Matthaeus, W.~H., \& Lamkin, S.~L. 1986, Physics of Fluids, 29, 2513,
  \dodoi{10.1063/1.866004}

\bibitem[{Passot {et~al.}(2017)Passot, Sulem, \& Tassi}]{passot2017}
Passot, T., Sulem, P.~L., \& Tassi, E. 2017, Journal of Plasma Physics, 83,
  715830402, \dodoi{10.1017/S0022377817000514}

\bibitem[{Passot {et~al.}(2018)Passot, Sulem, \& Tassi}]{Passot2018}
---. 2018, Physics of Plasmas, 25, 042107, \dodoi{10.1063/1.5022528}

\bibitem[{Ricci {et~al.}(2004)Ricci, Brackbill, Daughton, \&
  Lapenta}]{Ricci2004}
Ricci, P., Brackbill, J.~U., Daughton, W., \& Lapenta, G. 2004, Physics of
  Plasmas, 4102, \dodoi{10.1063/1.1768552}

\bibitem[{Roytershteyn \& Delzanno(2018)}]{Roytershteyn2018}
Roytershteyn, V., \& Delzanno, G.~L. 2018, Frontiers in Astronomy and Space
  Sciences, 5, 27, \dodoi{10.3389/fspas.2018.00027}

\bibitem[{{Sahraoui} {et~al.}(2013){Sahraoui}, {Huang}, {Belmont}, {Goldstein},
  {R{\'e}tino}, {Robert}, \& {De Patoul}}]{sahraoui13a}
{Sahraoui}, F., {Huang}, S.~Y., {Belmont}, G., {et~al.} 2013, The Astrophysical
  Journal, 777, 15, \dodoi{10.1088/0004-637X/777/1/15}

\bibitem[{Salem {et~al.}(2012)Salem, Howes, Sundkvist, Bale, Chaston, Chen, \&
  Mozer}]{Salem2012}
Salem, C.~S., Howes, G.~G., Sundkvist, D., {et~al.} 2012, Astrophysical Journal
  Letters, 745, 1, \dodoi{10.1088/2041-8205/745/1/L9}

\bibitem[{Stawarz {et~al.}(2016)Stawarz, Eriksson, Wilder, Ergun, Schwartz,
  Pouquet, Burch, Giles, Khotyaintsev, Contel, Lindqvist, Magnes, Pollock,
  Russell, Strangeway, Torbert, Avanov, Dorelli, Eastwood, Gershman, Goodrich,
  Malaspina, Marklund, Mirioni, \& Sturner}]{stawarz16}
Stawarz, J.~E., Eriksson, S., Wilder, F.~D., {et~al.} 2016, Journal of
  Geophysical Research, 121, 11021

\bibitem[{{Stver{\'a}k} {et~al.}(2015){Stver{\'a}k}, {Tr{\'a}vn{\'{\i}}{\v
  c}ek}, \& {Hellinger}}]{stverak2015}
{Stver{\'a}k}, {\AA}.~t., {Tr{\'a}vn{\'{\i}}{\v c}ek}, P.~M., \& {Hellinger},
  P. 2015, Journal of Geophysical Research (Space Physics), 120, 8177,
  \dodoi{10.1002/2015JA021368}

\bibitem[{TenBarge \& Howes(2013)}]{Tenbarge2013}
TenBarge, J., \& Howes, G. 2013, The Astrophysical Journall, 771, L27,
  \dodoi{10.1088/2041-8205/771/2/L27}

\bibitem[{{Treumann}(2009)}]{treumann09}
{Treumann}, R.~A. 2009, \aar, 17, 409, \dodoi{10.1007/s00159-009-0024-2}

\bibitem[{{{\v S}afr{\'a}nkov{\'a}} {et~al.}(2016){{\v S}afr{\'a}nkov{\'a}},
  {N{\v e}me{\v c}ek}, {N{\v e}mec}, {P{\v r}ech}, {Chen}, \&
  {Zastenker}}]{safrankova16}
{{\v S}afr{\'a}nkov{\'a}}, J., {N{\v e}me{\v c}ek}, Z., {N{\v e}mec}, F.,
  {et~al.} 2016, The Astrophysical Journal, 825, 121,
  \dodoi{10.3847/0004-637X/825/2/121}

\bibitem[{{{\v S}afr{\'a}nkov{\'a}} {et~al.}(2013){{\v S}afr{\'a}nkov{\'a}},
  {N{\v e}me{\v c}ek}, {P{\v r}ech}, \& {Zastenker}}]{safrankova13a}
{{\v S}afr{\'a}nkov{\'a}}, J., {N{\v e}me{\v c}ek}, Z., {P{\v r}ech}, L., \&
  {Zastenker}, G.~N. 2013, Physical Review Letters, 110, 025004,
  \dodoi{10.1103/PhysRevLett.110.025004}

\bibitem[{Vencels {et~al.}(2016)Vencels, Delzanno, Manzini, Markidis, Peng, \&
  Roytershteyn}]{Vencels2016}
Vencels, J., Delzanno, G.~L., Manzini, G., {et~al.} 2016, Journal of Physics:
  Conference Series, 719, 012022, \dodoi{10.1088/1742-6596/719/1/012022}

\bibitem[{Wan {et~al.}(2016)Wan, Matthaeus, Roytershteyn, Parashar, Wu, \&
  Karimabadi}]{Wan2016}
Wan, M., Matthaeus, W.~H., Roytershteyn, V., {et~al.} 2016, Physics of Plasmas,
  23, 042307, \dodoi{10.1063/1.4945631}

\bibitem[{Wan {et~al.}(2012)Wan, Matthaeus, Karimabadi, Roytershteyn, Shay, Wu,
  Daughton, Loring, \& Chapman}]{Wan2012}
Wan, M., Matthaeus, W., Karimabadi, H., {et~al.} 2012, Physical Review Letters,
  109, \dodoi{10.1103/PhysRevLett.109.195001}

\end{thebibliography}

\end{document}